\documentclass{sig-alternate-05-2015}

\usepackage{hyperref}
\usepackage{paralist}
\usepackage{caption}

\usepackage{graphicx}
\usepackage{amsmath}

\newtheorem{prop}{Proposition}[section]
\newtheorem{theorem}{Theorem}[section]
\usepackage{amssymb}
\usepackage{ifsym}
\usepackage{stmaryrd}
\usepackage{algorithm}
\usepackage{algorithm,algorithmicx,algpseudocode}
\renewcommand{\algorithmiccomment}[1]{\bgroup\hfill\scriptsize//~#1\egroup}
\usepackage{tikz}
\usetikzlibrary{arrows,positioning,automata,decorations,fit,backgrounds,calc,shapes,snakes}
\usetikzlibrary{shapes.geometric} 
\usetikzlibrary{decorations.pathmorphing} 
\usepackage{xcolor}
\usepackage{pgfplots}

\usepackage[compatibility=false]{caption}
\usepackage[labelformat=simple]{subcaption}

\newcommand{\OPT}{\mathbin{\mathrm{OPT}}}
\newcommand{\bound}{\mathbin{\mathrm{bound}}}
\newcommand{\UNION}{\mathbin{\mathrm{UNION}}}

\newcommand{\FILTER}{\mathbin{\mathrm{FILTER}}}
\newcommand{\SELECT}{\mathop{\mathrm{SELECT}}}

\newcommand{\true}{\mathit{true}}
\newcommand{\false}{\mathit{false}}
\newcommand{\error}{\mathit{error}}

\newcommand{\semm}[2]{\llbracket #1 \rrbracket_{#2}}

\newcommand{\dom}[1]{\mathrm{dom}(#1)}

\newcommand{\var}[1]{\mathit{vars}(#1)}

\newcommand{\AND}{\mathbin{\mathrm{AND}}}
\newcommand{\GRAPH}{\mathbin{\mathrm{GRAPH}}}
\newcommand{\PTIME}{\mathbin{\mathrm{PTIME}}}
\newcommand{\NP}{\mathbin{\mathrm{NP\text{-complete}}}}
\newcommand{\coNP}{\mathbin{\mathrm{coNP\text{-complete}}}}
\newcommand{\PSPACE}{\mathbin{\mathrm{PSPACE}}}

\newcommand{\isLiteral}{\mathbin{\mathrm{isLiteral}}}
\newcommand{\isBlank}{\mathbin{\mathrm{isBlank}}}
\newcommand{\isIRI}{\mathbin{\mathrm{isIRI}}}
\newcommand{\str}{\mathbin{\mathrm{str}}}
\newcommand{\lang}{\mathbin{\mathrm{lang}}}

\newcommand{\langMatches}{\mathbin{\mathrm{langMatches}}}
\newcommand{\Regex}{\mathbin{\mathrm{Regex}}}
\newcommand{\isChanged}{\mathbin{\mathrm{isChanged}}}
\def\sharedaffiliation{%
\end{tabular}
\begin{tabular}{c}}

\definecolor{blu1}{RGB}{102,140,250}
\definecolor{blu2}{RGB}{102,140,200}
\definecolor{blu3}{RGB}{102,140,150}

\tikzstyle{chart}=[
    legend label/.style={font={\scriptsize},anchor=west,align=left},
    legend box/.style={rectangle, draw, minimum size=5pt},
    axis/.style={black,semithick,->},
    axis label/.style={anchor=east,font={\tiny}},
]

\tikzstyle{pie chart}=[
    chart,
    slice/.style={line cap=round, line join=round, very thick,draw=white},
    pie title/.style={font={\bf}},
    slice type/.style 2 args={
        ##1/.style={fill=##2},
        values of ##1/.style={}
    }
]

\pgfdeclarelayer{background}
\pgfdeclarelayer{foreground}
\pgfsetlayers{background,main,foreground}

\newcommand{\pie}[3][]{
    \begin{scope}[#1]
    \pgfmathsetmacro{\curA}{90}
    \pgfmathsetmacro{\r}{1}
    \def\c{(0,0)}
    \node[pie title] at (90:1.3) {#2};
    \foreach \v/\s in{#3}{
        \pgfmathsetmacro{\deltaA}{\v/100*360}
        \pgfmathsetmacro{\nextA}{\curA + \deltaA}
        \pgfmathsetmacro{\midA}{(\curA+\nextA)/2}

        \path[slice,\s] \c
            -- +(\curA:\r)
            arc (\curA:\nextA:\r)
            -- cycle;
        \pgfmathsetmacro{\d}{max((\deltaA * -(.5/50) + 1) , .5)}

        \begin{pgfonlayer}{foreground}
        \path \c -- node[pos=\d,pie values,values of \s]{$\v\%$} +(\midA:\r);
        \end{pgfonlayer}

        \global\let\curA\nextA
    }
    \end{scope}
}

\begin{document}

\setcopyright{acmcopyright}

\doi{xxx.xxx/xxx}

\isbn{xxx-xxxx-xx-xxxx}



%

\title{On the Statistical Analysis of Practical SPARQL Queries}
\numberofauthors{6}
\author{
      \alignauthor Xingwang Han$^{1,2}$\\
      \email{xingwanghan@tju.edu.cn}
      \alignauthor Zhiyong Feng$^{1,2}$\\
      \email{zyfeng@tju.edu.cn}
      \alignauthor Xiaowang Zhang$^{1,2}$\\
      \email{xiaowangzhang@tju.edu.cn}
\and
      \alignauthor Xin Wang$^{1,2}$\\
      \email{wangx@tju.edu.cn}
      \alignauthor Guozheng Rao$^{1,2}$\\
      \email{rgz@tju.edu.cn}
      \alignauthor Shuo Jiang$^{3}$\\
      \email{jiangshuo@tju.edu.cn}
      \sharedaffiliation
      \affaddr{$^1$School of Computer Science and Technology, Tianjin University, Tianjin, China}  \\
      \affaddr{$^2$Tianjin Key Laboratory of Cognitive Computing and Application, Tianjin,
      China}\\
      \affaddr{$^3$School of Computer Software, Tianjin University, Tianjin, China}
}

\maketitle
\begin{abstract}
In this paper, we analyze some basic features of SPARQL queries coming from our practical world in a statistical way. These features include three statistic features such as the occurrence frequency of triple patterns, fragments, well-designed patterns and four semantic features such as monotonicity, non-monotonicity, weak monotonicity (old solutions are still served as parts of new solutions when some new triples are added) and satisfiability. All these features contribute to characterize SPARQL queries in different dimensions. We hope that this statistical analysis would provide some useful observation for researchers and engineers who are interested in what practical SPARQL queries look like, so that they could develop some practical heuristics for processing SPARQL queries and build SPARQL query processing engines and benchmarks. Besides, they can narrow the scope of their problems by avoiding those cases that do possibly not happen in our practical world.
\end{abstract}

%
%


%
%

%
%



\section{Introduction}\label{sec:intro}
The Resource Description Framework (RDF) \cite{RDFprimer}, firstly recommended by the World Wide Web Consortium (W3C) in 1998~\cite{perez_sparql_tods}, is a directed, labeled graph data format for representing information in the Semantic Web. RDF, as a graph model \cite{wood_survey,walklogic}, is often used to represent personal information, social networks, metadata about digital artifacts as well as to provide a means of integration over disparate sources of information. The SPARQL query language released by the RDF Data Access Working Group in 2004 becomes the official W3C Recommendation for RDF query language \cite{sparql} in 2008. It is an important language in graph databases which use graph structures with nodes, edges and properties to represent and store data \cite{ipl14,cj15}. SPARQL allows for a query consisting of triple patterns, conjunctions ($\AND$), disjunctions ($\UNION$), optional patterns ($\OPT$) and built-in conditions (constraints) to be filtered ($\FILTER$). The standard query language for RDF data is SPARQL \cite{sparql1.1}. Current version~1.1 of SPARQL extends SPARQL~1.0 \cite{sparql} with important features such as aggregation and regular expressions. Other features, such as negation and subqueries, have also been added, but mainly for efficiency reasons. They were already expressible by a more roundabout manner in version~1.0 (this follows from known results to the effect that every relational algebra query is expressible in SPARQL \cite{perez_sparql_tods,jair15}.). Hence, it is still relevant to study the fundamental properties of SPARQL~1.0.

In this paper, we present some statistics of seven basic features on practical SPARQL queries coming from our real world. In particular, we analyze a log of SPARQL queries, harvested from Linked SPARQL Query Log Dataset (LSQ) published in 2015: a
public, openly accessible dataset of SPARQL queries extracted from endpoint logs where the DBpedia SPARQL Endpoint is included \cite{saleem2015lsq}. The dataset contains more than 1.19 million unduplicated queries in total. These features include three statistic features, namely, \emph{the occurrence frequency of triple patterns}, \emph{fragments}, \emph{well-designed patterns} and four semantic features, namely, \emph{monotonicity}, \emph{non-monotonicity}, \emph{weak monotonicity} (old solutions are still served as parts of new solutions when some new triples are added) and \emph{satisfiability} which is used to determine whether a query is in error or meaningless. Though there are some existing works in statically analyzing SPARQL queries \cite{Francois,USEWOD2011,journals/pvldb/LetelierPPS12,pp_sparqlcontainment,DBLP:conf/ijcai/Guido15,DBLP:journals/pvldb/CebiricGM15,saleem2015lsq}, they either study some semantic features in a qualitative way \cite{USEWOD2011,journals/pvldb/LetelierPPS12,pp_sparqlcontainment,DBLP:conf/ijcai/Guido15,DBLP:journals/pvldb/CebiricGM15} or just analyze some statistic features \cite{Francois,saleem2015lsq}. As far as we known, it is still open to analyze these semantic features in a quantitative way. However, we have investigated that it is not trivial to do this since these semantic features of a single query might become totally different from semantic features of fragments. For instance, the fragment $AO$,  whose patterns contain only two operators $\AND$ and $\OPT$, is non-monotonic while a pattern $(?x, p, ?y)\ \OPT\ (?x, q, ?z)$ is weakly monotonic. Moreover, we have investigated that many queries can be determined  whether they are satisfiable or not, even they belong to fragments whose satisfiability is undecidable. For instance, a pattern $(?x, p, ?y)\ \FILTER \ (?x \neq ?z) \wedge (?y = c)$ belongs to the fragment SPARQL$(=, \neq)$ whose satisfiability is undecidable \cite{jair15}.

The main goal of this paper is to present a comprehensive statistical report about these seven features over SPARQL queries in LSQ. To simplify our discussion, we mainly consider these queries in SPARQL 1.0 (over 99.94\%). The main contributions of this paper can be summarized as follows:
\begin{compactitem}
\item We analyze three statistic features: the occurrence of triple patterns, fragments and well-designed patterns. And then we find the followings: a) over 96 \% of practical queries contain at most 7 triple patterns while those queries with at least 8 triple patterns are less than 4\%. b) Among 32 fragments of SPARQL 1.0, the four fragments: \emph{none}, A, F and AO, cover over 85\% of practical queries and the remaining 22 fragments contain less 15\% of practical queries. Additionally, the six fragments: FGO, FGU, GOU, FGOU, AFGU and AFGOU, do not occur in our practical world. c) These practical queries with well-designed patterns known to be weakly monotonic is about 77.66\%. About 22.30\% of practical queries are not well-designed patterns but still weakly monotonic.
\item We consider three semantic features: monotonicity, non-monotonicity and weak monotonicity. And then we find that among these practical queries, monotonic queries is 65.61\% and non-monotonic queries is about 0.04\%. In other words, practical queries are almost weakly monotonic (about 99.96\%). As a result, the evaluation of over 65.61\% queries is in PTIME and the evaluation of over 99.96\% queries is coNP-Complete while the evaluation with PSPACE-completeness involves in less 0.04\% practical queries.
\item We discuss an important semantic feature, namely satisfiability. We develop a sound algorithm to determine whether a given query with well-designed patterns is satisfiable or not. Finally, we can determine the satisfiability of all practical queries by analyzing common statistical structures of queries with non-monotonic well-designed patterns. As we excepted, all practical queries are almost satisfiable (over 99.99\%), which means the meaningless queries written by users are few in the real world.
\end{compactitem}

The rest of this paper is organized as follows: Section \ref{sec:pre} briefly
introduces SPARQL 1.0. Section \ref{sec:syntax} discusses three
statistic features. Section \ref{sec:monotone} and Section
\ref{sec:satisfiable} discuss four semantic features. Finally, Section
\ref{sec:con} summarizes the paper. Due to the limited space, we omit all
proofs but a full technique report with consisting of all proofs can be
found at a public website
\footnote{\url{http://123.56.79.184/Han2016TR.pdf}}.

\section{Syntax and semantics of SPARQL}\label{sec:pre}
In this section, we briefly recall the syntax and semantics of
SPARQL 1.0, largely depending on the excellent expositions
\cite{perez_sparql_tods,sparql}.

\noindent \textbf{RDF graphs} Let ${I}$, ${B}$ and ${L}$ be infinite sets of \emph{IRIs},
\emph{blank nodes} and \emph{literals}, respectively.  These
three sets are pairwise disjoint.  We denote the union $I \cup B
\cup L$ by $U$ and elements of $I \cup L$ will be referred to as
\emph{constants}. A triple $(s, p, o) \in ({I}\cup {B}) \times {I} \times ({I} \cup
{B}\cup {L})$ is called an \emph{RDF triple}. An \emph{RDF graph} is a finite set of RDF triples.

\noindent \textbf{Patterns} Assume furthermore an infinite set $V$ of \emph{variables}, disjoint from $U$. It is a SPARQL convention to prefix each variable with a question mark. Any triple from $({I}\cup {L} \cup {V}) \times ({I} \cup {V}) \times ({I} \cup {L} \cup {V}$) is a pattern (called a
\emph{triple pattern}). \emph{Patterns} are constructed by using triple patterns and operators $\UNION$, $\AND$, $\OPT$, $\GRAPH$ and $\FILTER$. Formally, patterns are of the forms as follows: $P_{1} \UNION P_{2}$, $P_{1} \AND P_{2}$, $P_{1} \OPT P_{2}$, $\GRAPH_i \ (P)$, $\GRAPH_{?x}\ (P)$ and $P \FILTER C$ where $i\in I$ and $C$ is a \emph{constraint}. A constraint is a boolean combination of the 17 atomic constraints.

\noindent \textbf{Semantics} The semantics of patterns are defined in terms of sets of so-called \emph{mappings}, which are simply total functions
$\mu \colon S \to U$ on some finite set $S$ of variables.
We denote the domain $S$ of $\mu$ by $\dom \mu$. Now given an RDF graph $G$ and a set of named graphs $\delta$, $D = (G, \delta)$ denotes a RDF dataset. Given a pattern $P$, we define the semantics
of $P$ on $D$, denoted by $\semm P D$, is a set of mappings whose satisfaction on constraints is based on a
three-valued logic with truth values $\true$, $\false$ and $\error$.

\noindent {\textbf{Queries}} A SELECT query is an expression of the form
$\SELECT_{S} P$ where $S$ is a finite set of variables and $P$ is
a pattern. Semantically, given an RDF dataset $D$, we define $\semm
{\SELECT_{S} P}{D} = \{\mu|_{\dom \mu \cap S} \mid \mu \in \semm
{P}{D}\}$, where we use the common notation $f|_X$ for the
restriction of a function $f$ to a subset $X$ of its domain.

\section{Statistic features of queries}\label{sec:syntax}
Our query log is extracted from LSQ \cite{saleem2015lsq}
which is a public dataset of SPARQL
queries extracted from the logs of public SPARQL endpoints.
It contains 5.7 million query executions.
LSQ is extracted from the following four SPARQL query logs:
DBpedia (from 30/04/2010 to 20/07/2010, 232 million triples), Linked Geo Data (LGD) (from 24/11/2010 to 06/07/2011: 1 billion triples), Semantic Web Dog Food (SWDF) (from 16/05/2014 to 12/11/2014: 300 thousand triples) and
British Museum (BM) (from 08/11/2014 to 01/ \\ 12/2014: 1.4 million triples).

We firstly collect 1,749,069 unduplicated queries from \\ 5,675,204 query executions. Secondly, we remove 555,084 queries which have parse error or do not follow the syntax of SPARQL 1.0. Indeed, it is still acceptable since only 0.6\% queries are beyond SPARQL 1.0. Finally, we collect 1,087,544 (91.5\% of SPARQL 1.0) SELECT queries by deleting all non-SELECT queries (i.e., ASK queries, DESCRIBE queries, or CONSTRUCT queries). So, in this paper, we mainly work on \textbf{1,087,544} queries from LSQ.

\subsection{Occurrence frequency of triple patterns}
The occurrence frequency of triple patterns is simply the number of triple patterns occurring in a query \cite{Francois}. As a basic feature, it directly influences the size of queries. In our investigation, we see the followings:
\begin{compactitem}
\item Over 54.99\% queries contain only one triple pattern in one query;
\item Most of practical queries contain at most 7 triple patterns (96\%);
\item The maximal number of triple patterns occurring in a query is 24.
\end{compactitem}

\begin{figure}[H]
\caption{Frequency of occurrences}\label{fig:tripleandfragment}
\centering
\scalebox{0.68}{
\begin{tikzpicture}
\begin{axis}[
    width=12cm,height=10cm,
    bar width=6pt,
    xbar,
    ylabel={Number of Triple Patterns},
    xlabel={Occurrence times},
    xmax = 650000,
    symbolic y coords={0,24,18,10,11,13,15,12,16,9,14,17,8,7,6,5,4,2,3,1},
    ytick=data,
    enlarge y limits=0.05,
	enlarge x limits=0,
    nodes near coords,
    point meta=x / 1087544, 
    nodes near coords align = {horizontal}
    ]
\addplot coordinates {(16,0) (152,24) (512,18) (544,10)
    (1359,11) (1790,13) (2282,15) (2349,12) (2493,16)
    (3161,9) (4334,14) (6482,17) (8302,8) (14610,7)
    (14902,6) (18546,5) (32681,4) (101074,2) (273884,3)
    (598071,1) };
\end{axis}
\end{tikzpicture}
}
\end{figure}
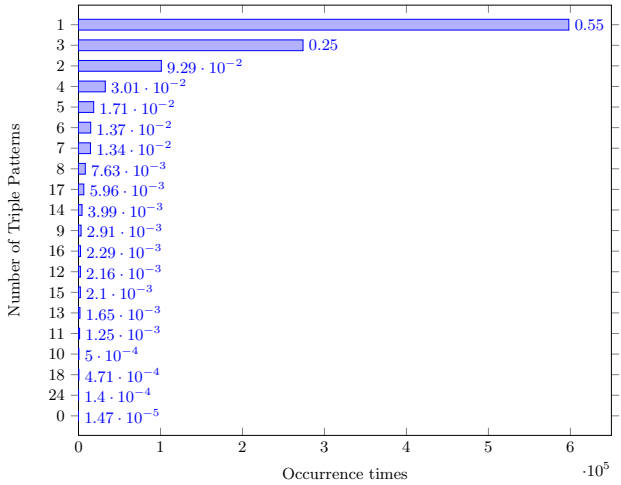

The full distribution of the occurrence frequency of triple patterns can be found in  Figure~\ref{fig:tripleandfragment}.

This statistical result shows that a practical query generally contains a few triple patterns.
It is helpful to restrict the scale of queries to be discussed.
Since most of queries have few triple patterns,
the shapes of graphs which are formed by the triple patterns are enumerable.
So we can propose an optimized scheme for the enumerable shapes in the future.

\subsection{Fragments}
Though the complexity of a query depends on its occurrence frequency of triple patterns, the evaluation also relies on its grammatical structure. Therefore, it is necessary to take the grammatical structure of queries into account. Now, we investigate which fragments of SPARQL are popular among practical queries \cite{saleem2015lsq}.

Let us abbreviate the operator $\AND$ by `A'; $\OPT$ by `O';
$\UNION$ by `U'; $\FILTER$ by `F'; and $\GRAPH$ by `G'. Then we can denote any fragment of SPARQL, where the letter word is formed by a subset of the five operators. We use `\emph{none}' to denote the fragment whose queries contain no operator.

We can find some interesting phenomena as follows:
\begin{compactitem}
\item Over 37\% queries are in the fragment of \emph{none} which is the biggest fragment among all 32 fragments of SPARQL.
\item Only 6 fragments: FGO, FGU, GOU, FGOU, AFGU and AFGOU, do not occur in our practical world. That is most of fragments are still useful.
\item Total of four fragments: \emph{none}, AO, F and A, are over 85.83\% and there are over 94.8\% if considering four fragments: AOU, FO, O, AFOU additionally.
\end{compactitem}

As is well known, fragments of SPARQL have variational complexity of query evaluation. For instance, the complexity of query evaluation of AF is $\NP$ even $\UNION$ is added \cite{perez_sparql_tods}. In particular, the complexity of query evaluation becomes $\PSPACE$-hard once $\OPT$ is added \cite{perez_sparql_tods}.

Thus, our statistical result could characterize the query evaluation of practical SPARQL queries precisely. For instance,
the three fragments: \emph{none}, A and F are over 62.83\% whose query evaluation is $\PTIME$. Moreover, it is interesting to optimize query evaluation of these fragments such as AO, F and A instead of the full SPARQL since they are popular in our practical world.

\subsection{Well-designed patterns}
Since $\OPT$ operator brings more complexity to query evaluation (generally, PSPACE-complete) \cite{schmidt_sparqloptim} and these fragments with $\OPT$ operator are over 31.97\% of the total. It is interesting to discuss some restricted form of patterns with $\OPT$ in a lower complexity. The \emph{well-designed} patterns \cite{perez_sparql_tods} have been identified as a well-behaved class of SPARQL patterns, with
properties similar to the conjunctive queries for relational
databases. Thus the query evaluation of well-designed patterns becomes $\coNP$ \cite{chile_sparql_pods}.
Let $P$ be a pattern and $C$ be a constraint, we use $var(P)$ to denote the set of variables occurring in $P$ and $var(C)$ to denote the set of variables occurring in $C$.

A pattern $Q$ is \emph{safe}
if for every subpattern $P\ \FILTER\ C$ of $Q$, it holds that $var(C) \subseteq var(P)$.
A $\UNION$-free pattern $P$ is \emph{well-designed}
if $P$ is safe and, for every subpattern $P^{'} =\ (P_{1}\ \OPT\ P_{2})$ of $P$
and for every variable $?x$ occurring in $P_{2}$,
the following condition holds: If $?x$ occurs both inside $P_{2}$ and outside $P^{'}$, then it also occurs in $P_{1}$.

A pattern is \emph{$\UNION$ Normal Form} (UNF, for short) if it is in the form of
$P_{1}\ \UNION \ P_{2}\ \UNION \ldots \UNION P_{n}$,
where each $P_{i}\ (1 \le i \le n)$ is $\UNION$-free.

A pattern in UNF is well designed if every $P_{i}(1 \le i \le n)$ is
a $\UNION$-free well-designed pattern. A pattern is \emph{non-well-designed} if it is not well-designed. Analogously, a query is \emph{(non-)well-designed} if its pattern is (non-)well-designed.
A query $\SELECT_{S}(P)$ is called \emph{well-designed} if $P$ is well-designed.

\noindent \textbf{Procedure of determining well-designed queries}
Now, we will present a procedure to determine whether a query is well-designed in three steps:
\vspace*{-3pt}
\begin{description}
\item[Step 1] If a pattern contains the $\GRAPH$ operator or \\
$\FILTER$ conditions which are not build-in condition, then it is not well-designed. In this case, there are 210,064/19.32\% queries.
\begin{compactitem}
\item If a pattern is not in the form of UNF
then it is not well-designed. In this case, there are 32,598/3.00\% queries.
\item If a pattern is a UNF pattern,
then turn to \textbf{Step 2}. In this case, there are 6,745/0.62\% queries.
\item If a pattern is a $\UNION$-free pattern, then turn to \textbf{Step 3}. In this case, there are 838,137/0.62\% queries.
\end{compactitem}
\vspace*{-5pt}
\item[Step 2] For every $\UNION$-free subpattern of a UNF pattern, using \textbf{Step 3}. If all the $\UNION$-free subpatterns are well-designed,
then the UNF pattern is well-designed (where there are \\ 6,745/0.62\% queries); otherwise, it is not well-designed.
\vspace*{-13pt}
\item[Step 3] If a pattern is a $\UNION$-free well-designed pattern, then it is well-designed (where there are 837,839/77.04\% queries); otherwise, it is not well-designed (where there are 298/0.03\% queries).
\end{description}
We can conclude the following results:
\begin{compactitem}
\item 844,584/77.66\% queries are well-designed and the remaining 242,960/22.34\% queries are not well-designed;
\item Over 3/4 practical queries are \emph{weak monotonicity} (\cite{ch}, defined in Section \ref{sec:monotone}) which provides a good support to well-designed patterns;
\item According to above result, we can find that over 62.83\% practical queries have query evaluation in $\PTIME$ while 14.83\% practical queries have query evaluation in \\$\NP$ and 22.34\% practical queries have query evaluation in (possible) $\PSPACE$-hard.
\end{compactitem}

\section{Monotonicity, weak monotonicity and non-monotonicity}\label{sec:monotone}
In the statistics of well-designed patterns (see Section \ref{sec:syntax}), we can find further some interesting phenomena as follows:
\begin{compactitem}
\item over 62.83\% practical queries with query evaluation in $\PTIME$ are monotonic;
\item about 14.83\% practical queries with query evaluation in $\NP$ are not monotonic but weakly monotonic (\cite{ch}, defined later);
\item about 22.34\% practical queries with query evaluation in (possible) $\PSPACE$-Complete are possibly neither monotonic nor weakly monotonic.
\end{compactitem}

In other words, three semantic features (i.e., monotonicity, weak monotonicity, non-monotonicity) connect with the complexity of query evaluation.

In this section, we look into the full distribution of monotonicity, weak monotonicity and non-monotonicity among our practical queries.

\noindent {\textbf{Monotonicity and weak monotonicity}}
For every two RDF graphs $G_{1}$, $G_{2}$ such that $G_1\subseteq G_2$, a pattern $P$ is said to be \emph{monotonic} if  it holds that $\semm P {G_1} \subseteq \semm P {G_2}$. A pattern $P$ is said to be \emph{non-monotonic} otherwise.
Let $\mu_1$ and $\mu_2$ be two mappings. $\mu_1$ is \emph{subsumed} in $\mu_2$ denoted by $\mu_1\sqsubseteq \mu_2$ if $\dom{\mu_1} \subseteq \dom{\mu_2}$ and $\mu_1(?x) = \mu_2(?x)$ for all $?x \in \dom{\mu_1}$. Let $\Omega_1$ and $\Omega_2$ be two sets of mappings. For any mapping $\mu_1 \in \Omega_1$, we use $\Omega_1 \sqsubseteq \Omega_2$ if  there exists some mapping $\mu_2 \in \Omega_2$ such that $\mu_1 \sqsubseteq \mu_2$.
For any two RDF graphs $G_{1}$ and $G_{2}$, a pattern $P$ is \emph{weakly monotonic} if  $G_{1}\subseteq G_{2}$ implies  $\semm P {G_1} \sqsubseteq \semm P {G_2}$. Note that all monotonic queries are weakly monotonic but not vice versa and all weakly monotonic queries are non-monotonic but not vice versa.

In the following of this paper, we mainly count weakly monotonic queries by excluding monotonic queries, as is well known, monotonic queries are weakly monotonic.

\noindent {\textbf{Determining monotonicity, weak monotonicity and non-monotonicity}}
Since $\GRAPH_{v} (P)$ and $P\ \FILTER\ C$ have the same monotonicity (weak monotonicity or non-monotonicity) as $P$ \cite{chile_sparql_pods}, we can ignore the $\GRAPH$ operator and the difference between the non-build-in condition and build-in condition.
Moreover, queries in $\OPT$-free fragments are monotonic. Thus we can exclude these queries.

\noindent {\textbf{Main procedure}}
The main procedure contains the following steps:
\vspace*{-3pt}
\begin{description}
\item[Step 1] For every pattern, if it is not in UNF, we need to rewrite it to its UNF according to \cite[Lemma 3]{chile_sparql_pods}(where there are 44,101/4.06\%).
\vspace*{-5pt}
\item[Step 2] For every pattern in UNF, if all the $\UNION$-free sub-patterns in it are $\OPT$-free or $\OPT$-monotonic then it is monotonic. In this case, there are \\ 713,532/65.61\% queries. Otherwise, if all the $\UNION$-free sub-patterns are $\UNION$-free well-designed sub-patterns then the pattern is weakly monotonic \cite{chile_sparql_pods}. In this case, there are 330,389/30.38\% queries.
\end{description}

\vspace*{-3pt}
A $\UNION$-free pattern $P$ is $\OPT$-\emph{monotonic} if
$\var{P_{2}} \subseteq \var{P_{1}}$ for every subpattern $Q = P_{1}\ \OPT\ P_{2}$ of $P$.

Immediately, we can conclude the following proposition.
\begin{prop}\label{prop:opt-monontone}
Let $P$ be a $\UNION$-free pattern. If $P$ is $\OPT$-monotonic then $P$ is monotonic.
\end{prop}
\noindent {\textbf{Exceptions of determining procedure}}
In the first step of main procedure, not all patterns can be logically translated into their $\UNION$ normal form  due to the distributive law on $P_1\ \OPT\  (P_2 \ \UNION \ P_3)$ disabled (about 20,361/1.87\% queries),
where there are four cases: (1) $(?x, p, ?y) \OPT ((?x, q, ?z) \UNION (?x, r, ?u) \UNION (?x, s, ?v))$ \\ (18,466/1.70\%) (weakly monotonic); (2)
$(?x, p, a) \OPT ((b,\\ q, ?y) \UNION (c, r, ?y))$ (1,494/0.14\%) (weakly  monotonic); (3) $(?x, p, a) \OPT ((?x, q, ?y) \UNION (?z, r, b))$ (388/0.04\%) (weakly monotonic);
and (4) $((?x, p, q) \OPT (?y, q, a)) \OPT ((?y, q, b) \UNION (?z, r, c))$ (13/0.00\%) (non-monotonic).

Finally, among these queries which can be rewritten into their $\UNION$ normal forms, there still exist 23,262/2.14\% SPARQL queries that are neither well-designed nor can be rewritten to well-designed queries.

The remaining unknown SPARQL queries contain the following five cases:
(1) $((?x, p, a) \OPT (?x, q, ?y)) \FILTER \ (\text{langMatches}(\lang(?y), 'en'))$ (22,848/2.10\%) (montone);
(2) $(() \OPT (?x, p, a))  \OPT (?x, q, b)$ (77/0.01\%) (non-monotonic);
(3) $((?x, a, b) \OPT (?x, c, ?y)) \FILTER \neg \bound(?y)$ (213/0.02\%\\) (non-monotonic);
(4) $((?x, a, b) \OPT (?y, c, d)) \OPT (?y, d, e)$ (120/0.01\%) (non-monotonic); and
(5) $((?x, p, ?y) \OPT \\(?y, q, ?z)) \FILTER (\neg \bound(?z)) \vee (?x = a)$ (1/0.00\%) (non-monotonic).

At last, there are three SPARQL queries which are not safe. Since the queries are not safe, they are not satisfiable \cite{jair15} (discussed in Section~\ref{sec:satisfiable}). So, by default, these queries are monotonic.

\noindent {\textbf{Statistical results}} We can show the result in the following:
\begin{compactitem} 
\item It is a surprise that over 713,548/65.61\% queries are monotonic and about 373,578/34.35\% queries are weakly monotonic while only 418/0.04\% queries are non-monotonic. In other words, over 99.96\% queries are weakly monotonic.
\item In the weakly monotonic fragment(1,087,126 queries), there are 844,584/77.69\% queries are well-designed patterns. Moreover, there are about 242,542/22.31\% weakly monotonic queries which are not well-designed patterns.
\item In the non-well-designed fragment(242,996 queries), ov-er 222,194/91.45\% queries which are not well-designed can be logically translated into equivalent well-designed patterns and about 20,348/8.38\% queries which are  neither well-designed patterns nor can be rewritten to well-designed patterns but are still weakly monotonic. Otherwise, there are 418/0.17\% queries which are non-monotonic.
\end{compactitem}

%
%
%
%


\section{Satisfiability of queries}\label{sec:satisfiable}
Although the satisfiable problem of well-designed patterns is decidable, this  problem of the full SPARQL is undecidable \cite{jair15}. Base on this, we are interested to know how many queries are satisfiable.
Note that a query is called \emph{satisfiable} if there exists an RDF graph under which the pattern evaluates to a nonempty set of mappings.

\noindent {\textbf{Procedure of determining satisfiability}}
The main procedure of determining whether a query is satisfiable consists of six steps as follows:
\begin{description}
\item[Step 1] If $\semm P G \neq \emptyset$ for some RDF graphs in existing dataset then $P$ must be satisfiable, otherwise turn to \textbf{Step 2}. In this pre-processing step, we can process \\ 635,704/58.46\% queries.
\item[Step 2] If $P$ is a filter-free pattern then it is satisfiable \cite{jair15}, otherwise turn to \textbf{Step 3}. In this step, we can handle further 372753/34.28\% queries.
\item[Step 3] If $P$ contains negated bound constraint then put it into a pool to be determined at last (about 84 queries); otherwise enter \textbf{Step 4}.
\vspace*{-22pt}
\item[Step 4] If $P$ is well-designed or can be rewritten to a well-designed pattern then turn to \textbf{Step 5}; otherwise again put it into a pool to be determined last (about 13,630/1.25\% queries).
\vspace*{-22pt}
\item[Step 5] Using Algorithm~\ref{alg:satisfiable} to determine its satisfiability of the remaining queries (about 65,373/6.01\%).
\end{description}

\vspace*{-15pt}
Since the satisfiability of a well-designed pattern is equivalent to an $\OPT$-free pattern \cite[Proposition 1]{jair15}, we mainly consider the satisfiability of $\OPT$-free patterns in the following of this section. To determine the satisfiability of well-designed AF-queries, we need the following three sub-steps of \textbf{Step 6}:

\vspace*{-20pt}
\begin{description}
\item[Step 6.1] Rewrite constraints exhaustively by applying all inference rules in Table \ref{tab:inferenceRules}.
\vspace*{-25pt}
\item[Step 6.2] Translate all patterns into its \emph{strong $\UNION$ normal form} (strong UNF, for short) where a pattern is of the form $Q_1 \ \UNION \ldots \UNION \ Q_m$, where each $Q_i$ ($1\le i \le m$) is an AF-pattern and all constraints occurring in $Q_i$ are atomic.
\vspace*{-25pt}
\item[Step 6.3] Determining the closeness of the closure (i.e., collection) which includes all atomic constraints. If it is close then return ``unsatisfiable ''; otherwise return ``satisfiable''. A set of constraints is \emph{close}
if it subsumes a \emph{conflict} of the form $\{\alpha, \neg \alpha\}$ for an atomic constraint $\alpha$.
\end{description}

\vspace*{-50pt}
\begin{prop}\label{prop:inference-rules}
Let $C$ be a constraint. For any pattern $P$, $P\ \FILTER\ C \equiv P\
\FILTER\ C'$ where $C'$ is obtained from $C$ by using inference rules in
Table \ref{tab:inferenceRules}. Note that, all the atomic constraints can
infer itself, we omit the inference from it to itself.
\end{prop}

\vspace*{-20pt}
By Proposition \ref{prop:inference-rules}, we can conclude that all $\OPT$-free patterns can be logically translated into its strong $\UNION$ normal form.
\vspace*{-18pt}
\begin{prop}\label{prop:strong-UNF}
Let $P$ be an $\OPT$-free pattern. There exists some $Q$ in strong UNF such that $P \equiv Q$.
\end{prop}

\vspace*{-12pt}
Now, we conclude that Algorithm \ref{alg:satisfiable} is sound and complete.

\vspace*{-18pt}
\begin{theorem}\label{thm:sound-satisfiable}
Let $P$ be an AF-pattern. If $P$ is in strong $\UNION$ normal form then $P$ is satisfiable iff the closure of $P$ obtained in Algorithm \ref{alg:satisfiable} is not close.
\end{theorem}

\vspace*{-20pt}
At last, the remaining queries (about 13,630/1.25\%), which are neither well-designed patterns nor
able to be logically translated to equivalent well-designed queries, contains the following cases: (1) $((?x, a, b) \OPT (?y, c, d)) \FILTER \neg \bound(?y)$ (84/0.01\%) (unsatisfiable); (2)
$(P_{1} \OPT P_{2} \OPT ... \OPT P_{n})$ (13,488/(1.24\%)) (as the same as $P_1$); and (3) $(?x, a, b) \\\FILTER \bound(?y)$ (58/0.00\%) (satisfiable).

As we excepted, over 99.99\% of practical queries are satisfiable, that is, the meaningless queries written by users are few in our practical world.

Zhang and Van den bussche\cite{jair15} presented a determination method of satisfiability for well-designed patterns. In this paper, we implement an algorithm to determine the satisfiability of practical queries.

\begin{algorithm}[H]
	\caption{Determining the satisfiability of well-designed pattern}
	\label{alg:satisfiable}
	\begin{algorithmic}[1]
		\Require Well-designed pattern $P$ in strong $\UNION$ normal form
		\Ensure Determining the satisfiability of $P$
        \State $P = (Q_{1} \UNION Q_{2} \UNION ... \UNION Q_{m})$
		\For {every $Q_{j}$($1 \leq j \leq m$) in $P$}
            \State $S = \{C_{1}, C_{2}, ... , C_{k}\}$, ($1 \leq j \leq k, C_{j}$ is constraint in $Q_{j}$)
            \State $\mathcal{L} = \{S\}$
            \Repeat
                \For {every $S_{i}$ in $\mathcal{L}$}
                    \If {$\isChanged_{i} = \false$}
                        \State continue;
                    \Else
                        \State $\isChanged_{i} = \false$
                    \EndIf
                    \For {every $C_{j}$ in $S_{i}$}
                        \If {$C_{j} \longrightarrow D \text{ and } D \notin S_{i}$}
                            \State $S_{i} = S_{i} \cup D$, $isChanged_{i} = \true$
                        \EndIf
                    \EndFor
                    \If {$C_{j} = C_{1} \land C_{2}$}
                        \If {$\{C_{1}, C_{2}\} \not \subseteq S_{i}$}
                            \State $S_{i} = S_{i} \cup \{C_{1}, C_{2}\}$, $\isChanged_{i} = \true$
                        \EndIf
                    \EndIf
                    \If {$C_{j} = C_{1} \lor C_{2}$}
                        \If {$\{C_{1}, C_{2}\} \cap S_{i} = \O$}
                            \State $S_{i} = S_{i} \cup \{C_{1}\}$, $S_{i}^{'} = S_{i} \cup \{C_{2}\}$, $\mathcal{L} = \mathcal{L} \cup \{S_{i}^{'}\}$, $\isChanged_{i} = \true$
                        \EndIf
                    \EndIf
                \EndFor
            \Until {every $\isChanged_{i}$ is $\false$}

            \State $\mathcal{L} = \{S_{1}, S_{2}, ... ,S_{l}\}$
            \If {there exists $S_{i}$($1 \leq i \leq l$) is consistent}
                \State \Return $P$ is satisfiable.
            \EndIf
        \EndFor
        \State \Return $P$ is unsatisfiable.
	\end{algorithmic}
\end{algorithm}

\section{Conclusions}\label{sec:con}
In this paper, we have presented comprehensive statistics of seven basic features of a log of SPARQL 1.0 queries in LSQ. We think that these statistical results could provide some useful observation for researchers and engineers who are interested in what practical {SPARQL} queries look like. In the future, we are going to analyze other interesting features of SPARQL 1.0 queries.

\newpage
\thispagestyle{empty}
\pagestyle{empty}

\begin{table}[t]
\caption{Inference Rules}
\label{tab:inferenceRules}
$$
\begin{array}[t]{|c|c|c}
\hline
\textbf{Atomic} & \textbf{Inferred} \\ \textbf{constraints} & \textbf{constraints} \\
\hline
?x=c & \bound(?x)\\
\hline
?x \neq c & \bound(?x)\\
\hline
?x=?y & \bound(?x) \land \bound(?y) \\
\hline
?x \neq ?y & \bound(?x) \land \bound(?y)\\
 \hline
?x>c & ((?x \geq c) \land (?x \neq c)) \land \isLiteral(?x) \\ & \land (\neg \isBlank(?x) \land \neg \isIRI(?x)) \\ & \land \bound(?x)\\
 \hline
 ?x \le c & ((?x < c) \lor (?x = c)) \land \isLiteral(?x)\\  &  \land (\neg \isBlank(?x)) \land \neg \isIRI(?x)) \\ & \land \bound(?x)\\
  \hline
?x < c & ((?x \leq c) \land (?x \neq c)) \land \isLiteral(?x) \\  & \land (\neg \isBlank(?x) \land \neg \isIRI(?x)) \\ & \land \bound(?x)\\
 \hline
?x \ge c & \land ((?x > c) \lor (?x = c)) \land \isLiteral(?x)\\ & \land (\neg \isBlank(?x) \land \neg \isIRI(?x)) \\ & \land \bound(?x)\\
 \hline
\isLiteral(?x) & (\neg \isBlank(?x) \land \neg \isIRI(?x)) \\& \land \bound(?x)\\
 \hline
\neg \isLiteral(?x) & (\isBlank(?x) \lor \isIRI(?x)) \\ & \land \bound(?x)\\
 \hline
\isBlank(?x) & (\neg \isLiteral(?x) \land \neg \isIRI(?x)) \\ & \land \bound(?x)\\
\hline
\neg \isBlank(?x) & (\isLiteral(?x) \lor \isIRI(?x)) \\ & \land \bound(?x)\\
\hline
\isIRI(?x) & (\neg \isBlank(?x) \land \neg \isLiteral(?x)) \\ & \land \bound(?x)\\
\hline
\neg \isIRI(?x) & (\isLiteral(?x) \lor \isBlank(?x)) \\& \land \bound(?x)\\
\hline
\str(?x)=c & (\isLiteral(?x) \lor \isIRI(?x)) \\
 & \land\neg \isBlank(?x) \land \bound(?x) \\
 \hline
\str(?x) \neq c& (\isLiteral(?x) \lor \isIRI(?x)) \\
 & \land \neg \isBlank(?x) \land \bound(?x) \\
 \hline
\lang(?x)=c & \langMatches(\lang(?x), c) \land \\ & \isLiteral(?x) \land (\neg \isBlank(?x) \land \\ & \neg \isIRI(?x)) \land \bound(?x) \\
 \hline
\lang(?x) \neq c& \neg \langMatches(\lang(?x), c) \land \\ & \isLiteral(?x) \land (\neg \isBlank(?x) \land \\ & \neg \isIRI(?x)) \land \bound(?x) \\
 \hline
\langMatches & \lang(?x)=c \land \isLiteral(?x) \land \\
(\lang(?x), c) & (\neg \isBlank(?x) \land \neg \isIRI(?x)) \\ & \land \bound(?x)\\
 \hline
\neg \langMatches& (\lang(?x) \neq c) \land \isLiteral(?x) \land \\
 (\lang(?x), c) & (\neg \isBlank(?x) \land \neg \isIRI(?x)) \\ & \land \bound(?x) \\
 \hline
\Regex & (\isLiteral(?x) \lor \isIRI(?x)) \land \\ (\str(?x), r)  & \neg \isBlank(?x) \land \bound(?x)\\
 \hline
\neg \Regex & (\isLiteral(?x) \lor \isIRI(?x)) \\ (\str(?x), r)
 & \land \bound(?x) \land \neg \isBlank(?x)
  \\
 \hline
\Regex(?x, r) & \Regex(\str(?x), r) \land \isLiteral(?x) \\ & \land (\neg \isBlank(?x) \land \isIRI(?x)) \\ & \land \bound(?x)\\
 \hline
\neg \Regex(?x, r) & (\neg \Regex(\str(?x), r) \land \isLiteral(?x) \\
& \land (\neg \isBlank(?x) \land \neg \isIRI(?x)) \\ & \land \bound(?x)\\
 \hline
\end{array}
$$
\end{table}

\end{document}